\documentstyle[aps,prl,epsfig,multicol]{revtex}

\catcode`\@=11
\def\preprint#1{%
\def\@preprint{\noindent\flushright{#1}\vskip 10pt}%
}

\def\beq{\begin{equation}}
\def\eeq{\end{equation}}
\def\beqa{\begin{eqnarray}}
\def\eeqa{\end{eqnarray}}

\begin{document}

\newcommand\pt{p_{\rm \scriptscriptstyle T}}
\newcommand\hatpt{\hat{p}_{\rm \scriptscriptstyle T}}

\preprint{Bicocca-FT-02-5\\UPRF-2002-4}

\title{Is There a Significant Excess in Bottom Hadroproduction 
at the Tevatron?}
\author{Matteo Cacciari}
\address{Dipartimento di Fisica,
Universit\`a di Parma, Italy, and\\
INFN, Sezione di Milano, Gruppo Collegato di Parma}
\author{Paolo Nason}
\address{INFN, Sezione di Milano\\
Via Celoria 16, 20133 Milan, Italy}

\maketitle

\begin{abstract}
We discuss the excess in the hadroproduction of $B$ mesons
at the Tevatron. We show that an accurate use of up-to-date
information on the $B$ fragmentation function reduces
the observed excess to an acceptable level. Possible implications for
experimental results reporting  \emph{bottom quark}
cross sections, also showing an excess with respect to 
next-to-leading order theoretical predictions, are discussed.
\end{abstract}

\begin{multicols}{2}
 
Since a few years, bottom production
has been one of the very few instances in which experimental results
and Quantum Chromodynamics (QCD) predictions have sometimes displayed not 
too good an agreement.
Bottom quark hadroproduction cross sections have been measured by the
UA1 Collaboration~\cite{Albajar:1987iu,Albajar:1991zu} at the CERN $Sp\bar pS$ collider and by
both the CDF~\cite{Abe:1993sj,Abe:1993hr,Abe:1995dv}
and D0~\cite{Abbott:1999se} experiments at the Fermilab
Tevatron in $p\bar p$ collisions, and found to be about a factor of two
or more larger than  next-to-leading order (NLO) QCD
predictions \cite{Nason:1988xz,Nason:1989zy,Beenakker:1991ma}.
CDF has recently published data for $B^+$ meson
production~\cite{Acosta:2001rz}. They claim
an excess over QCD predictions by about a
factor of three. 
The H1 and ZEUS experiments at the electron-proton collider HERA have both
measured  $D^*$ production cross sections \cite{Breitweg:1998yt,Adloff:1998vb}.
Both measurements are compatible with QCD calculations
\cite{Frixione:2002zv}, although the ZEUS data is on the high side
of the theoretical uncertainty band.
Bottom production has been measured at HERA and from photon-photon collisions
at LEP, and found to be larger than predictions, by about a factor of
three or more
\cite{Acciarri:2000kd,OPALNote:2000sc,Adloff:1999nr,Breitweg:2000nz}. 
The cross
section for the production of the heaviest quark, the top, has also
been measured at the Tevatron, and found instead perfectly compatible
with theoretical expectations.

By pushing the parameters of the theoretical calculation to somewhat extreme
values, it is not impossible to accommodate the bottom spectrum
observed at the Tevatron.
Alternatively, one can take the discrepancy more seriously,
and invoke some ``new physics'' contribution~\cite{Berger:2000mp}
in order to explain it.
It has also been known for a long time that the fixed order, NLO
QCD calculation may be insufficient to explain the data,
because of the presence of some enhanced contributions, that
can be included via resummation of large classes of Feynman diagrams,
and that contribute positively to the cross sections.
These contributions are threshold effects, small-$x$ effects,
and high transverse momentum logarithms, which may be important since
most of the cross section is measured at large transverse momentum.
A full calculation of
next-to-next-to-leading QCD contributions, years ahead in the future,
might finally also contribute to explain the apparent discrepancy.

In this Letter we shall not try to improve on the perturbative aspects of heavy
quark production. We shall instead focus our attention on
a specific non-perturbative issue, namely on the implementation
of hadronization effects.
In fact, we shall argue that a good part of the discrepancy between
theory and data arises
when one tries to supplement the perturbative
prediction for $b$ quark production with a non-perturbative model
for the formation of a $B$ meson from the $b$ quark,
or, alternatively, to correct the data in an attempt to give
a $b$ quark spectrum rather than a $B$ meson one.
The non-perturbative hadron formation effect is usually introduced by
writing the hadron-level cross section for $B$ mesons as
\beq
\frac{d\sigma^B}{d\pt}
= \int d\hatpt dz \frac{d\sigma^b}{d\hatpt}
D(z) \;\delta(\pt-z\hatpt)\;,
\label{Bhad}
\eeq
the function $D(z)$ being a phenomenological parametrization of hadronization
effects.
Traditionally, the Peterson et al.~\cite{Peterson:1983ak} $D(z;\epsilon)$ 
form of the fragmentation function is used,
implemented in conjunction with a quark cross section given by
a shower Monte Carlo program. The $\epsilon$
parameter is obtained from fits to $e^+e^-$ data \cite{Chrin:1987yd}.
The effect of fragmentation is to
reduce the momentum of the $B$ meson with respect to that of the $b$ quark. 
It is roughly a 10\% effect, being of the order of $\overline{\Lambda}/m$, 
where $\overline{\Lambda}$ is a hadronic scale, of the order of a few
hundred MeV, and $m$ is the bottom quark mass.
It has however an important impact on the value of the
cross section, because of the steeply falling
transverse momentum spectrum of the $b$ quark. Since transverse momentum cuts
are always applied, the measurable cross section is strongly reduced
by this effect. It should be clear from this discussion that,
in order to assess the presence of a discrepancy in the $B$ production
data, the effect of fragmentation should be assessed clearly and
unambiguously.

In Ref.~\cite{Acosta:2001rz}
the CDF Collaboration compares its data to a theoretical
prediction obtained by convoluting the NLO cross section for bottom
quarks with a Peterson fragmentation function.
They use $\epsilon=0.006\pm 0.002$, which is the traditional
value proposed in Ref.~\cite{Chrin:1987yd}. They claim that their
data is a factor of 2.9 higher than the QCD calculation.

The purpose of this Letter is precisely to implement correctly
the effect of heavy quark fragmentation in the QCD calculation.
Several ingredients are necessary in order to do this:
\begin{itemize}
\item A calculation with resummation of large transverse momentum
logarithms at the next-to-leading level (NLL) should be
used for heavy quark production \cite{Cacciari:1994mq}, in order to
correctly account for scaling violation in the fragmentation function.
\item A formalism for merging the NLL resummed results with
the NLO fixed order calculation (FO) should be used,
in order to account properly
for mass effects \cite{Cacciari:1998it}. This calculation will be
called FONLL in the following.
\item A NLL formalism should be used to extract the non-perturbative 
fragmentation
effects from $e^+e^-$ data~\cite{Mele:1991cw,Colangelo:1992kh,%
Dokshitzer:1996ev,%
Cacciari:1997wr,Cacciari:1997du,Nason:1999zj,Cacciari:2001cw}.

\end{itemize}
We begin by pointing out that, as shown in
Refs.~\cite{Cacciari:1997du,Nason:1999zj}, the value $\epsilon=0.006$
is appropriate only when a leading-log (LL) calculation of the spectrum 
is used, as
is the case in shower Monte Carlo programs. When NLL calculations are
used, smaller values of $\epsilon$ are needed to fit the data.
It must further be pointed out that, as noted
in~\cite{Frixione:1998ma,Nason:1999ta}, 
it is not the detailed knowledge of the whole
spectrum of $D(z)$ in $z\in [0,1]$ to be relevant for the
calculation of hadronic cross sections.
For the steeply falling differential distributions $d\sigma/d\pt$,
that have usually a power law behaviour, the knowledge of some specific moment
of the fragmentation function
\begin{equation}
D_N \equiv\int D(z) z^N \frac{dz}{z}
\end{equation}
is sufficient to obtain the hadronic cross section. In fact, assuming
that $d\hat{\sigma}/d\hatpt = A\hatpt^{-n}$
in the neighborhood of some $\hatpt$ value, one immediately finds
\begin{equation}
\frac{d\sigma}{d\pt} =\int dz d\hatpt\, D(z) \frac{A}{\hatpt^n}
 \,\delta(\pt-z \hatpt) = \frac{A}{\pt^n} D_n\;.
\end{equation}
Thus, the hadronic cross section is given by the product of the partonic
cross section times the $n^{\rm th}$ moment of the fragmentation function,
where $n$ is the power behaviour of the cross section in the neighborhood
of the value of $\pt$ being considered. In Ref.~\cite{Nason:1999ta} it
is also shown that this is an excellent approximation to the exact integral
in the cases of interest.
The value of $n$ for the $\pt$ spectrum in the region
of interest ranges from 3 to 5.
It is therefore clear that, when fitting $e^+e^-$ data, getting a good
determination of the {\sl moments} of the non-perturbative fragmentation
function between 3 and 5 is more important than attempting to
describe the whole $z$ spectrum.

\begin{center}
\begin{figure}[ht]
\epsfig{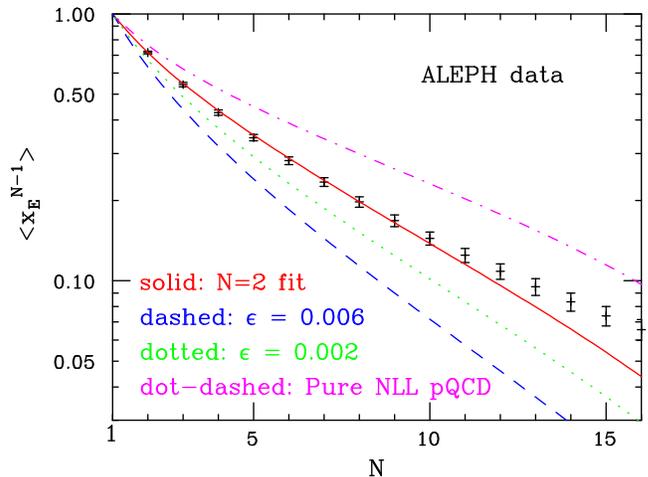}
\caption{
Moments of the measured $B$ meson fragmentation function, 
compared with the perturbative
NLL calculation supplemented with different $D(z)$ non-perturbative 
fragmentation forms.
The solid line is obtained using a one-parameter form fitted to the
second moment.
}
\label{fig:moments}
\end{figure}
\end{center}

Fig.~\ref{fig:moments} shows the moments calculated from the $x_E$ (the 
$B$ meson energy fraction with respect to the beam energy)
distribution data for weakly decaying $B$ mesons in $e^+e^-$ collisions 
published by
the ALEPH Collaboration~\cite{Heister:2001jg}. The experimental error
bars shown in the plot have been evaluated by taking into account the full
bin-to-bin correlation matrix \cite{Boccali:2002}. Four curves are
superimposed to the data. All of them have been obtained with an
underlying NLL perturbative description~\cite{Mele:1991cw,Cacciari:2001cw}.
The bottom quark mass $m$ has been taken equal to 4.75 GeV and the QCD
scale has been fixed to $\Lambda^{(5)} = 0.226$ GeV. Sudakov resummation has not
been included, since its effect is negligible in the low-moment
region~\cite{Cacciari:2001cw}. These are the
default values of the parameters that we shall use in this work
for the computation of the hadronic cross section.

The dot-dashed line represents the purely perturbative part.
The dashed line represents the convolution of the perturbative part
described above with a Peterson form with $\epsilon = 0.006$. It is
evident that this produces a poor description of even the lowest
moments. The dotted line is obtained using $\epsilon = 0.002$, a value known to
produce good fits of the $x_E$ distribution when used together with a NLL
perturbative calculation~\cite{Cacciari:1997du,Nason:1999zj}. The
description of the moments improves, but the line still cannot fall
within the error bars. There is thus a problem in obtaining a good fit
of the low moments of the fragmentation function using the Peterson
parametrization. The problem can be traced back to the need to fit
points with very large $x_E$ (where most of the $e^+e^-$ data is) since
there the perturbative calculation becomes less reliable.
Normally, the very large $x_E$ region is excluded from the fit because
of this reason. The computed cross section is thus allowed to become
negative in this region, a fact that leads to an underestimate
of the low moments.

It should be clear from the aforementioned arguments that,
in order to make accurate predictions for hadronic cross sections,
the non-perturbative part of the fragmentation should be fitted
in such a way that the low moments are well reproduced.
This is shown in the solid line in the figure.
A one-parameter form of the non-perturbative fragmentation
function has been used\footnote{Exactly which
functional form is used is actually not relevant. However, for the
sake of completeness, we mention that we have used a normalized
Kartvelishvili et al. form~\cite{Kartvelishvili:1978pi}, $D(z;\alpha)
= (\alpha+1)(\alpha+2) z^\alpha (1-z)$.}  and its free parameter has
been fixed by fitting the $N=2$ point in moments space, i.e. the
average energy fraction $\langle x_E\rangle$.  In this case the
functional form is good enough to describe well the
experimental data up to $N\simeq 10$. It is therefore a good candidate
to be employed in the calculation of the hadronic cross section
according to Eq.~(\ref{Bhad}). We shall refer to this fit in the
following as the ``$N=2$ fit''.

We notice that the effect of non-perturbative fragmentation
(i.e., the ratio of the dashed, dotted and solid curves with the dot-dashed
curve) is considerably reduced when the moments are fitted. Thus, perturbation 
theory alone
gives a much better description of the low moments of the fragmentation
function, rather than of its shape in $x$-space, requiring less non-perturbative 
input.
This is a consequence of the fact that, at large $x$, many
enhanced non-perturbative contributions and hadronization effects due
to the limited phase space come into play. Indeed, the form of the leading power
correction in moments space is well known, and reads~\cite{Nason:1997pk}
\begin{equation}
D_N = 1 - (N-1)\frac{\overline{\Lambda}}{m} + {\cal
O}\left(\frac{{\overline{\Lambda}}^2}{m^2}\right)\;.
\end{equation}
It can easily be checked that the form we
employed in the $N=2$ fit is consistent with this leading power correction,
provided one replaces $\alpha$ with $2m/\overline{\Lambda}$. 
One can also clearly see the non-perturbative 
correction to be minimal for $N=2$.
It is therefore always
desirable to study moments, rather than the $x$ shape of the fragmentation
function, also in order to perform QCD studies.

\begin{figure}[th]
\begin{center}
\epsfig{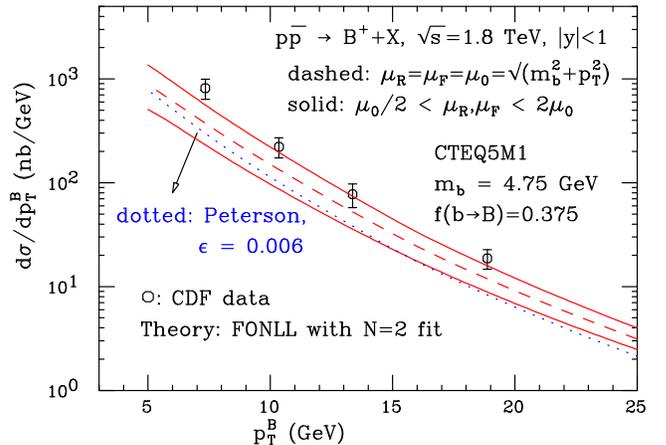}
\caption{Prediction for the $B$ cross section, obtained using the
calculation of Ref.~\protect\cite{Cacciari:1998it} supplemented with the
$N=2$ fit of the non-perturbative fragmentation function,
compared to the CDF data
of Ref.~\protect\cite{Acosta:2001rz}. For comparison, the result obtained using
a Peterson form with $\epsilon=0.006$ is also shown.}
\label{fig:Bhad}
\end{center}
\end{figure}

\begin{figure}[th]
\begin{center}
\epsfig{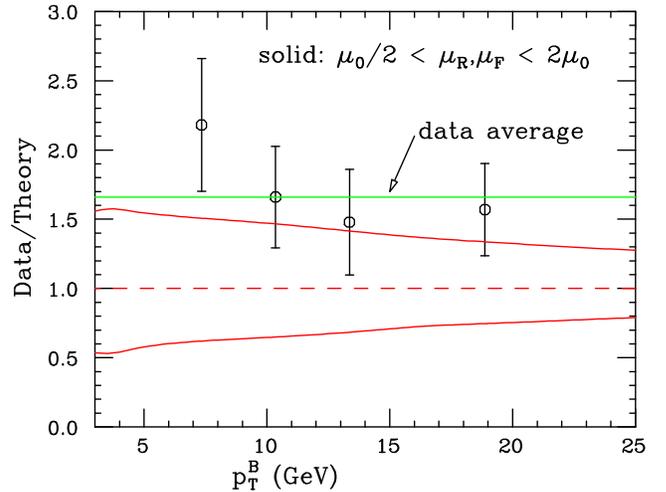}
\caption{Data over Theory ratio for $B$ production. Data points and theoretical
curves are as in Fig.~\ref{fig:Bhad}}
\label{fig:Bhadratio}
\end{center}
\end{figure}

In Figs.~\ref{fig:Bhad} and \ref{fig:Bhadratio} we show the final
prediction for $B$ hadroproduction at the Tevatron, 
obtained by the procedure outlined above. It is clearly shown how the
``Peterson with 
$\epsilon=0.006$'' choice underestimates the $B$ cross section at
large values of $\pt$. It is also clear that the claimed discrepancy
of a factor of 2.9 is now reduced to a factor of 1.7 with respect to
the central value prediction. The band obtained by varying the scales
gives an idea of the theoretical error involved, and, as one can see, the data
are not far above it.

\begin{figure}[ht]
\begin{center}
\epsfig{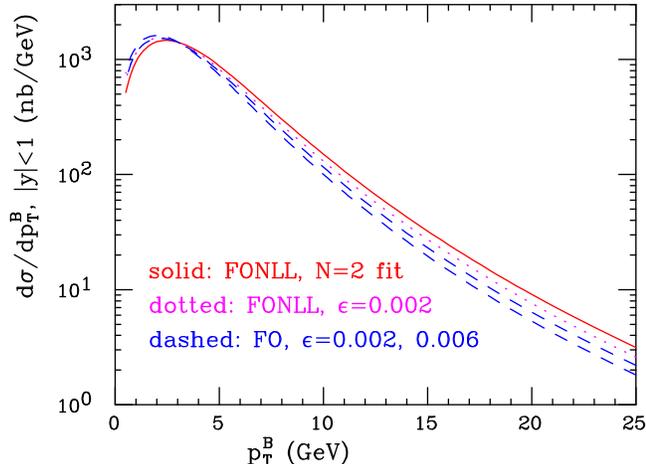}
\caption{
The effect of the different ingredients in the calculation presented
in this work, relative to a fixed order calculation with Peterson fragmentation
and $\epsilon=0.006$.}
\label{fig:breakup}
\end{center}
\end{figure}

At this point, we wish to quantify to what extent the various ingredients
of the present calculation affect the computed cross section, so
that it is in fact larger than the one given in Ref.~\cite{Acosta:2001rz}.
This is shown in Fig.~\ref{fig:breakup}.
From the figure we see that the FO (Fixed Order) calculation
with $\epsilon=0.006$ is the lowest curve. Using the more appropriate
value $\epsilon=0.002$ brings about a 20\%{} increase of the cross
section at $\pt=20$.
Using the FONLL calculation of
Ref.~\cite{Cacciari:1998it} brings about another 20\%{} increase,
and so does also the use of the $N=2$ fit.
The total effect is an increase by a factor of $1.2^3\simeq 1.7$, which
turns the factor of 2.9 reported in Ref.~\cite{Acosta:2001rz}
into the $1.7$ observed here.

Our FO, $\epsilon=0.006$ cross section is also higher
than the one presented in Ref.~\cite{Acosta:2001rz} in the low $\pt$
region. This difference could be due to the different possible
treatments of fragmentation at small transverse momentum. We have applied the
fragmentation to the momentum, rather than the energy or the $+$
component, of the fragmenting particle. We believe that these other choices,
although acceptable in the large-$\pt$ region, are not appropriate
in the non-relativistic limit.

The SLD experiment has also published accurate data on the $b$
fragmentation function \cite{Abe:2002iq}. Using their data instead
of the ALEPH ones brings about a slight decrease of the cross section,
below 4\%\ in the region of interest. On the other hand, using
more recent parton distribution function sets \cite{Pumplin:2002vw}, 
we find an increase of the
predicted cross section between 4 and 8\%\ in the region of interest.

In summary, we find that an appropriate treatment of the fragmentation
properties of the $b$ quark considerably reduces the discrepancy of the
CDF transverse momentum spectrum for the $B$ mesons and the corresponding
QCD calculation. The experimental points are compatible with predictions
obtained using the
present value of the QCD scale parameter and of the structure functions,
and a $b$ pole mass of $4.75$~GeV, and lie near the upper region
of the theoretical band obtained by varying the factorization and
renormalization scales. Including experimental and theoretical
uncertainties, the updated Data/Theory ratio can be written as 1.7 $\pm$ 0.5
(expt) $\pm$ 0.5 (theory).
 The calculation we have adopted includes
in a consistent way fixed order QCD results and the NLL resummation
of transverse momentum logarithms. Furthermore, the ``moments'' method
introduced 
here avoids the difficult large-$x$ region in the fragmentation function,
that would require more complex treatment and introduce further
uncertainties. 

While we have here convoluted a perturbative prediction to get a hadron level
result, the opposite path is followed by experiments when they deconvolute 
their hadron level cross sections in order to publish quark level
data~\cite{Albajar:1987iu,Albajar:1991zu,Abe:1993sj,Abe:1993hr,Abe:1995dv,Abbott:1999se}. 
In the light of what argued in this Letter, and of the apparent excess also 
shown by those 
data, it will be advisable to investigate whether a similar bias might have
affected those results.

In the meantime, we emphasize the importance of the
direct measurements of the moments of the fragmentation function
for heavy quarks. Measuring directly moments instead of the $x$ distribution
could be useful for the purpose of QCD
studies, and also for the computation of production
cross sections in hadronic collisions.

\acknowledgements
We wish to thank Tommaso Boccali for useful conversation.
MC wishes to thank Mario Greco for the invitation to 
the La Thuile conference, which prompted this investigation, and, more
importantly, for the extensive collaboration on this subject.


\begin{thebibliography}{10}

\bibitem{Albajar:1987iu}
C. Albajar {\it et~al.}, Phys. Lett. {\bf B186},  237  (1987).

\bibitem{Albajar:1991zu}
C. Albajar {\it et~al.}, Phys. Lett. {\bf B256},  121  (1991).

\bibitem{Abe:1993sj}
F. Abe {\it et~al.}, Phys. Rev. Lett. {\bf 71},  500  (1993).

\bibitem{Abe:1993hr}
F. Abe {\it et~al.}, Phys. Rev. Lett. {\bf 71},  2396  (1993).

\bibitem{Abe:1995dv}
F. Abe {\it et~al.}, Phys. Rev. Lett. {\bf 75},  1451  (1995).

\bibitem{Abbott:1999se}
B. Abbott {\it et~al.}, Phys. Lett. {\bf B487},  264  (2000).

\bibitem{Nason:1988xz}
P. Nason, S. Dawson, and R.~K. Ellis, Nucl. Phys. {\bf B303},  607  (1988).

\bibitem{Nason:1989zy}
P. Nason, S. Dawson, and R.~K. Ellis, Nucl. Phys. {\bf B327},  49  (1989),
erratum-ibid.\ B {\bf 335} (1989) 260.

\bibitem{Beenakker:1991ma}
W. Beenakker {\it et~al.}, Nucl. Phys. {\bf B351},  507  (1991).

\bibitem{Acosta:2001rz}
D. Acosta {\it et~al.}, Phys. Rev. {\bf D65},  052005  (2002).

\bibitem{Breitweg:1998yt}
J. Breitweg {\it et~al.}, Eur. Phys. J. {\bf C6},  67  (1999).

\bibitem{Adloff:1998vb}
C. Adloff {\it et~al.}, Nucl. Phys. {\bf B545},  21  (1999).

\bibitem{Frixione:2002zv}
S. Frixione and P. Nason, JHEP {\bf 03}, 053 (2002).

\bibitem{Acciarri:2000kd}
M. Acciarri {\it et~al.}, Phys. Lett. {\bf B503},  10  (2001).

\bibitem{OPALNote:2000sc}
OPAL Physics Note PN455, August 29 2000.

\bibitem{Adloff:1999nr}
C. Adloff {\it et~al.}, Phys. Lett. {\bf B467},  156  (1999), erratum-ibid.
  {\bf B 518}, 331-332, (2001).

\bibitem{Breitweg:2000nz}
J. Breitweg {\it et~al.}, Eur. Phys. J. {\bf C18},  625  (2001).

\bibitem{Berger:2000mp}
E.~L. Berger {\it et~al.}, Phys. Rev. Lett. {\bf 86},  4231  (2001).

\bibitem{Peterson:1983ak}
C. Peterson, D. Schlatter, I. Schmitt, and P.~M. Zerwas, Phys. Rev. {\bf D27},
  105  (1983).

\bibitem{Chrin:1987yd}
J. Chrin, Z. Phys. {\bf C36},  163  (1987).

\bibitem{Cacciari:1994mq}
M. Cacciari and M. Greco, Nucl. Phys. {\bf B421},  530  (1994).

\bibitem{Cacciari:1998it}
M. Cacciari, M. Greco, and P. Nason, JHEP {\bf 05},  007  (1998).

\bibitem{Mele:1991cw}
B. Mele and P. Nason, Nucl. Phys. {\bf B361},  626  (1991).

\bibitem{Colangelo:1992kh}
G. Colangelo and P. Nason, Phys. Lett. {\bf B285},  167  (1992).

\bibitem{Dokshitzer:1996ev}
Y.~L.~Dokshitzer, V.~A.~Khoze and S.~I.~Troian,
Phys. Rev. {\bf D53}, 89 (1996).
%

\bibitem{Cacciari:1997wr}
M. Cacciari, M. Greco, S.~Rolli, and A.~Tanzini, Phys. Rev. {\bf D55},  2736
  (1997).

\bibitem{Cacciari:1997du}
M. Cacciari and M. Greco, Phys. Rev. {\bf D55},  7134  (1997).

\bibitem{Nason:1999zj}
P. Nason and C. Oleari, Nucl. Phys. {\bf B565},  245  (2000).

\bibitem{Cacciari:2001cw}
M. Cacciari and S. Catani, Nucl. Phys. {\bf B617},  253  (2001).

\bibitem{Frixione:1998ma}
S. Frixione, M.~L. Mangano, P. Nason, and G. Ridolfi,
``Heavy Flavors II'', eds. A.J. Buras and M. Lindner,
 Adv. Ser. Direct. High
  Energy Phys. {\bf 15},  609  (1998).

\bibitem{Nason:1999ta}
P. Nason {\it et~al.},
Report of the 1999 CERN Workshop on SM physics (and more) and the LHC,
hep-ph/0003142.

\bibitem{Heister:2001jg}
A. Heister {\it et~al.}, Phys. Lett. {\bf B512},  30  (2001).

\bibitem{Boccali:2002}
T. Boccali, private communication.

\bibitem{Kartvelishvili:1978pi}
V.~G. Kartvelishvili, A.~K. Likhoded, and V.~A. Petrov, Phys. Lett. {\bf B78},
  615  (1978).

\bibitem{Nason:1997pk}
P. Nason and B.~R. Webber, Phys. Lett. {\bf B395},  355  (1997).

\bibitem{Abe:2002iq}
K. Abe {\it et~al.}, SLD Coll., SLAC-PUB-9087, Feb 2002, hep-ex/0202031.

\bibitem{Pumplin:2002vw}
J. Pumplin {\it et~al.}, CTEQ Coll., MSU-HEP-011101, Jan 2002, hep-ph/0201195.

\end{thebibliography}

 \end{multicols}
\end{document}